%% file: main.tex
\documentclass[conference]{IEEEtran}
\IEEEoverridecommandlockouts
\usepackage{cite}
\usepackage{amsmath,amssymb,amsfonts}
\usepackage{algorithmic}
\usepackage{graphicx}
\usepackage{textcomp}
\usepackage{xcolor}
\usepackage{comment}
\usepackage{url}
\usepackage[nolist,nohyperlinks]{acronym}

\usepackage[normalem]{ulem}

\usepackage{pgfplots}
\usetikzlibrary{patterns}
\pgfplotsset{compat=newest}

\def\BibTeX{{\rm B\kern-.05em{\sc i\kern-.025em b}\kern-.08em
    T\kern-.1667em\lower.7ex\hbox{E}\kern-.125emX}}

\newcommand{\red}[1]{{\textcolor[rgb]{1,0,0}{#1}}}

\renewcommand{\figurename}{Figure}

\begin{document}
\title{Performance Analysis of Tri-Sector Reflector Antennas for HAPS-Based Cellular Networks}

\author{
\IEEEauthorblockN{
German Svistunov\IEEEauthorrefmark{1},
Matteo Bernabè\IEEEauthorrefmark{1},
and
David López-Pérez\IEEEauthorrefmark{1}\IEEEauthorrefmark{2}
}
\normalsize\IEEEauthorblockA{\emph{
\IEEEauthorrefmark{1}Universitat Politècnica de València (UPV), Spain} 
}
\normalsize\IEEEauthorblockA{\emph{
\IEEEauthorrefmark{2}Beihang Valencia Polytechnic Institute (BVPI), China} 
}
\thanks{
979-8-3315-6128-4/26/\$31.00 ©2026 IEEE. Personal use of this material is permitted. Permission from IEEE must be obtained for all other uses, in any current or future media, including reprinting/republishing this material for advertising or promotional purposes, creating new collective works, for resale or redistribution to servers or lists, or reuse of any copyrighted component of this work in other works.

This work has been accepted for publication in the 7th International Conference on Communications, Signal Processing, and their Applications (ICCSPA 2026), to be held 15-18 June 2026 in Alcalá de Henares (Madrid), Spain.

This research is supported by the Generalitat Valenciana, Spain, through the CIDEGENT PlaGenT, Grant CIDEXG/2022/17, Project iTENTE, and the action CNS2023-144333, financed by MCIN/AEI/10.13039/501100011033 and the European Union “NextGenerationEU”/PRTR. }

}

\maketitle

\input{Sections/00_Abstract}
\input{Sections/01_Introduction}

\input{Sections/01a_Related_Work}

\input{Sections/02_System_model}

\input{Sections/03_Results}
\input{Sections/04_Conclusion}
\bibliographystyle{ieeetr}
\bibliography{journalAbbreviations, refs}

\begin{acronym}[AAAAAAAAAAAAAAAAAAAAAAAA]  
 \acro{3GPP}{third generation partnership project}
 \acro{4G}{fourth-generation}
 \acro{5G}{fifth-generation}
 \acro{6G}{sixth-generation}
 \acro{BS}{base station}
 \acro{CoMP}{coordinated multi-point}
 \acro{CSI-RS}{channel state Information reference signal}
 \acro{dB}{decibel}
 \acro{dBi}{decibel isotropic}
 \acro{DFT}{discrete Fourier transform}
 \acro{FR}{Frequency Range}
 \acro{FSO}{free-space optical}
 \acro{HAPS}{high-altitude platform station}
 \acro{HPBW}{half-power beamwidth}
 \acro{IoT}{internet of things}
 \acro{IRS}{intelligent reflecting surfaces}
 \acro{ISD}{inter-site distance}
 \acro{LoS}{line-of-sight}
 \acro{LTE}{long term evolution}
 \acro{mMIMO}{massive Multiple Input/Multiple Output}
 \acro{mmWave}{millimetre wave}
 \acro{NLoS}{non-line-of-sight}
 \acro{NR}{new radio}
 \acro{NTN}{non-terrestrial network}
 \acro{O2I}{outdoor-to-indoor}
 \acro{PRB}{physical resource block}
 \acro{RF}{radio frequency}
 \acro{SINR}{signal-to-interference-plus-noise ratio}
 \acro{SNR}{signal to noise ratio}
 \acro{SSB}{synchronization signal block}
 \acro{TBS}{terrestrial base station}
 \acro{THz}{terahertz}
 \acro{TN}{terrestrial network}
 \acro{UAV}{unmanned aerial vehicle}
 \acro{UE}{user equipment}
 \acro{UMa}{urban macro}
 \acro{UPA}{uniform planar array}
 \acro{MIMO}{multiple-input multiple-output}
 \acro{RSRP}{reference signal received power}
\end{acronym}

\end{document}

%% file: Sections/00_Abstract.tex
\begin{abstract}
The increasing demand for ubiquitous, high-capacity mobile connectivity has driven cellular systems to explore beyond-terrestrial deployments. In this paper, we present a system-level performance evaluation of fifth-generation (5G) non-terrestrial network enabled by high-altitude platform station (HAPS)-based base stations equipped with tri-sector reflector antennas against fourth-generation (4G) terrestrial network (TN) and 5G TN deployments in a multi-cell dense urban environment. Using the simulation results comprising the average effective downlink signal-to-interference-plus-noise ratio and the average user throughput, along with the subsequent interference analysis, we demonstrate that the reflector-based HAPS architecture is primarily constrained by inter-cell interference, while the combination of reflector configuration and deployment altitude represents a key design parameter.
\end{abstract}




%% file: Sections/01_Introduction.tex
\section{Introduction}

The growing demand for ubiquitous wireless connectivity has driven the evolution of cellular networks beyond traditional terrestrial infrastructures.
While \ac{5G} \acp{TN} have achieved significant advances in spectral efficiency, data rates, and latency, their coverage remains fundamentally constrained by ground-based \ac{BS} deployment density and terrain limitations. 
To address these challenges, \acp{NTN} have emerged as a promising complement to terrestrial systems, enabling seamless coverage from the sky through satellite, aerial, and high-altitude platforms. 
Among these technologies, 
\ac{HAPS} represent a particularly attractive solution,
providing a trade-off between the wide coverage of satellites and the low latency of terrestrial networks, while enabling flexible integration between \ac{TN} and \ac{NTN} architectures \cite{svistunov2025bridgingearthspacesurvey}.

The integration of \ac{HAPS} into the \ac{5G} and beyond framework has been actively pursued within the \ac{3GPP} \ac{NTN} standardization process,
where detailed channel models, antenna configurations, and deployment scenarios have been analyzed~\cite{3GPP38811}. 
However, despite the growing interest in \ac{HAPS}-based \acp{NTN}, comprehensive system-level studies that directly compare their performance with conventional terrestrial cellular networks under realistic multi-cell deployments remain scarce. 
As a result, the potential of \ac{HAPS} systems to support dense traffic environments and act as a large-scale access infrastructure is still not well understood.
In particular, it remains unclear whether \ac{HAPS}-based deployments can deliver performance comparable to conventional terrestrial cellular networks in dense multi-cell environments.

Recent studies have demonstrated that the unique characteristics of \ac{HAPS} enable a wide range of applications that are not limited to coverage extension in remote, rural, and disaster-affected regions, 
but also include handling unexpected data traffic events,  \ac{IoT}, edge computing, \ac{UAV} connectivity and control, network energy efficiency optimization, and other emerging use cases~\cite{svistunov2025bridgingearthspacesurvey}. 
Considering the extensive growth of wireless subscriptions and traffic,
particularly with wearable devices, electric vehicles, and drones, 
the combination of fiber, terrestrial wireless, and non-terrestrial technologies will be crucial to meet broadband connectivity demands~\cite{ericsson2025reportNov}.
While high-capacity \acp{TN} are typically available in dense urban environments, they face limitations related to densification inefficiency, power consumption, and deployment complexity. Consequently, large-scale \ac{HAPS} deployment may be particularly beneficial for applications such as drone control, public safety, and disaster early warning~\cite{HAPSAlliance2026WhitePaper}.
These developments motivate the expansion of studies on \ac{HAPS}-based \ac{NTN} performance to urban scenarios.
In this context, 
this paper investigates a different paradigm in which \ac{HAPS}-based \acp{NTN} are evaluated as a potential large-scale access infrastructure capable of complementing or even replacing terrestrial cellular deployments in dense environments.
Specifically, we provide a comprehensive simulation-based performance evaluation of a \ac{HAPS}-based deployment compared with conventional \ac{4G} and \ac{5G} terrestrial networks under a dense urban multi-cell scenario.

The rest of this paper is organized as follows. 
Section~\ref{sec:RelatedWork} reviews related work on \ac{HAPS}-based \ac{NTN} design and prior comparative analyses.
Section~\ref{sec:SystemModel} describes the system model adopted in this work. Section~\ref{sec:Results} presents and discusses the obtained results.
Finally, Section~\ref{sec:Conclusion} concludes the paper and discusses directions for future research.


%% file: Sections/01a_Related_Work.tex
\section{Related Work}\label{sec:RelatedWork}

The integration of \ac{HAPS} into \acp{NTN} has garnered significant attention in recent literature.
In this section,
we review key contributions in performance evaluation, integration architectures, and specific \ac{HAPS} optimizations.


Anicho~\textit{et al.}~\cite{anicho2021multihaps} analyzed \ac{HAPS}-specific challenges in multi-\ac{HAPS} networks within the \ac{3GPP} \ac{NTN} framework, 
including platform mobility, station-keeping for quasi-stationary positioning, and inter-\ac{HAPS} link design.
Xing~\textit{et al.}~\cite{xing2021architecture} investigated the system-level performance of fixed-wing \ac{HAPS}-based networks using both regenerative and transparent architectures.
The authors evaluate the energy efficiency and spectral efficiency of both configurations for single-cell (60 km radius) and multi-cell (100 km radius) scenarios under \ac{4G} \ac{LTE} Band 1 operation.
Results show that both architectures achieved comparable downlink and uplink spectral efficiencies, with the uplink limited by \ac{UE} transmit power and antenna gain.
Directional antennas at the \ac{UE} side significantly enhanced system throughput.
Although the authors provided comprehensive comparative analyses of \ac{HAPS} architectures, 
this work, however, is limited by rural scenarios,
and does not consider variations in \ac{HAPS} deployment altitude.
Lee~\textit{et al.}~\cite{lee2023performance} examined the downlink \ac{SINR}-based coverage performance of multiple \ac{HAPS} platforms using the antenna and channel models suggested by \ac{3GPP}.
However, the system model is limited by the small number of UEs and their highly specific spatial positioning,
while the performance evaluation focuses on \ac{SINR} and does not consider spectral efficiency, data rates, or latency.

Liu~\textit{et al.}~\cite{liu2022interference} studied inter-system downlink interference between coexisting \ac{TN} and \ac{HAPS}-based \ac{NTN} systems. They proposed dynamic interference coordination schemes which outperformed existing methods based on fixed resource allocation.
\ac{HAPS}-terrestrial network integration was also examined by Shamsabadi, Yadav, and Yanikomeroglu~\cite{shamsabadi2024enchancing} with a particular focus on the maximization of minimal spectral efficiency of the network through joint optimization of \ac{UE} association and beamforming weights. 
The proposed algorithm outperforms both standalone \ac{TN} scenarios and beamforming-only optimization with max-SINR association in terms of spectral efficiency.

Beyond performance evaluation and coexistence studies, several works have focused on the optimization of \ac{HAPS}-based cellular system design.
Shibata~\textit{et al.}~\cite{shibata2020system} presented a comprehensive optimization framework for designing multi-cell, gigabit-capable \ac{HAPS} systems that support \ac{4G} \ac{LTE} and \ac{5G} \ac{NR} \acp{UE}. 
Using genetic algorithms, the authors optimize cell configuration, antenna beamwidth, and tilt for various numbers of cells (7 to 21) to maximize throughput, while satisfying downlink and uplink \ac{SINR} constraints. 
In the study, energy efficiency and power consumption trade-offs are evaluated, accounting for \ac{HAPS} payload and power limitations. 
It was shown that multi-layer circular cell arrangements combined with uplink \ac{CoMP} reception techniques significantly enhance link reliability and throughput, enabling energy-efficient gigabit-class service from a single HAPS platform.
In their following paper~\cite{shibata2024haps}, 
Shibata~\textit{et al.} also considered coexistence and coverage optimization for HAPS-based networks operating alongside terrestrial mobile systems.
They proposed a cell design method that adapts \ac{HAPS} beam configurations and frequency reuse patterns to minimize interference and maximize service continuity. 
The model incorporates cell size optimization, antenna pattern control, and handover management, ensuring compatibility with existing terrestrial \ac{4G} \ac{LTE} infrastructures.
Simulation results confirmed that carefully optimized \ac{HAPS} cells can extend coverage beyond terrestrial limits (especially in rural and mountainous areas) while maintaining acceptable \ac{SINR} and spectral efficiency. 
The work highlights the importance of interference-aware cell geometry and the integration of \ac{HAPS} into hybrid \ac{TN}-\ac{NTN} ecosystems.


\subsection{Our contribution}

Overall, while existing studies provide valuable foundational insights into the integration of \ac{HAPS} within \acp{NTN},
they often lack detailed system-level evaluations that directly compare \ac{HAPS}-based deployments with conventional \acp{TN} in realistic multi-cell scenarios. 
Most prior works focus primarily on remote or rural environments, 
where \ac{HAPS}-based \acp{NTN} are envisioned as complementary systems intended to extend coverage or provide connectivity in sparsely populated regions where terrestrial infrastructure is economically or technically challenging to deploy. 
Consequently, in much of the literature, 
\ac{HAPS} platforms are treated as an auxiliary layer that supports existing \acp{TN},
rather than as a potential alternative to terrestrial cellular infrastructure. 
Furthermore, many studies adopt simplified system assumptions,
such as a limited number of \acp{UE}, idealized user distributions, or isolated single-cell configurations. 
In addition, the considered deployment configurations are often constrained to the conventional operational altitude range of 20--50\,km for \ac{HAPS},
without exploring how variations in platform altitude, antenna characteristics, and multi-cell interference conditions jointly affect network performance.

In contrast, this paper investigates a different paradigm in which \ac{HAPS}-based \acp{NTN} are considered as a potential primary connectivity layer capable of supporting dense urban deployments.
This perspective departs from the conventional view of \ac{HAPS} systems as complementary infrastructure and instead explores their potential to operate as a large-scale alternative to terrestrial cellular deployments.
Specifically, we present a comprehensive simulation-based performance evaluation of a \ac{HAPS}-based \acp{BS} deployment equipped with tri-sector reflector antennas operating in a dense urban multi-cell environment. 
The analysis focuses on downlink performance and evaluates key system-level metrics,
including the average effective \ac{SINR} experienced at the \ac{UE} side and the corresponding average \ac{UE} throughput. 
By examining the interplay between platform altitude and reflector antenna aperture, 
the study identifies the operational region in which \ac{HAPS}-based deployments can achieve their best performance. 
In addition, we analyze the inter-cell interference patterns that emerge in dense \ac{HAPS}-based \acp{BS} deployments. 
These results provide insights into the feasibility of \ac{HAPS}-based networks as a large-scale access infrastructure and help quantify the conditions under which such systems could complement or potentially replace conventional terrestrial cellular deployments.

%% file: Sections/02_System_model.tex
\section{System model}\label{sec:SystemModel}

\begin{figure}[tb]
\centerline{\includegraphics[width=0.45\textwidth]{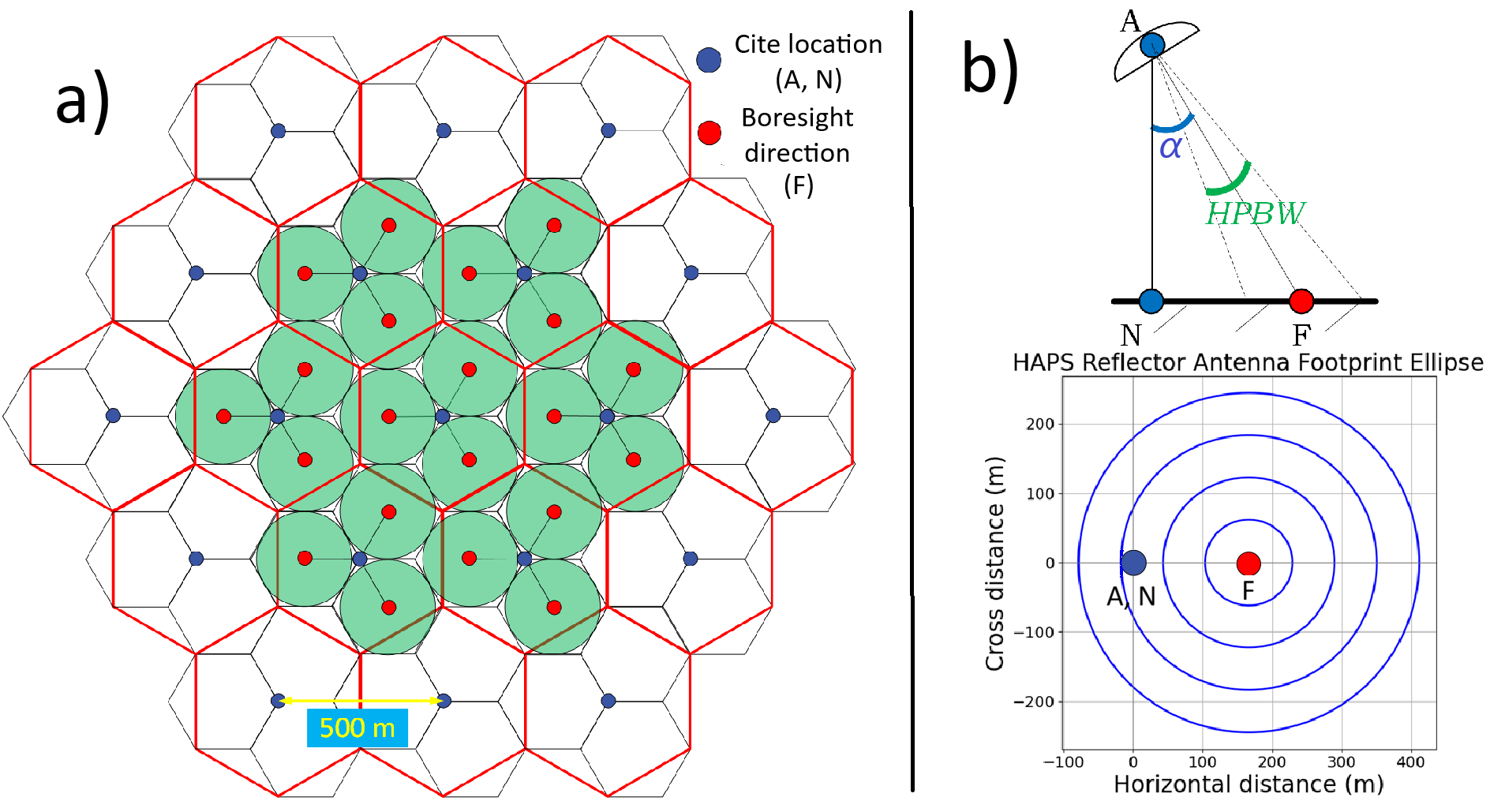}}
\caption{NTN deployment scenario over a hexagonal grid (a), reflector antenna tilt adjustment and HPBW footprint increase for different altitudes (1\,km, 2\,km, 3\,km, and 4\,km for the aperture radius of 10 wavelengths) (b).}
\label{fig:haps_deployment}
\end{figure}
In this section, we describe the system model adopted in this paper. We consider the downlink performance of terrestrial and non-terrestrial cellular deployments in an \ac{UMa} scenario, following the models outlined in~\cite{3GPP38811,3GPP38901}.

\subsection{Deployment scenarios}
Here, 
we describe the considered network deployment, operating frequency bands, and \ac{UE} distribution used in the simulations.
\begin{table}[!t]
    \centering
    \caption{Simulation parameters}
    \begin{tabular}{ccc}
        \textbf{Parameter} & \textbf{Notation} & \textbf{Reference values} \\
        4G Carrier frequency [GHz] & \( f_c^{\rm 4G} \) & 2 \\
        5G Carrier frequency [GHz] & \( f_c^{\rm 5G} \) & 3.5 \\
        4G Bandwidth [MHz] & \( B_0^{\rm 4G} \) & 20 \\
        5G Bandwidth [MHz] & \( B_0^{\rm 5G} \) & 100 \\
        Reflector aperture radius [wavelengths] & \(r_{\lambda}\) & 5..50 \\
        Reflector aperture radius [m] & \(r_{m}\) & 0.43..4.3 \\
        HAPS altitude [km] & \( h_{\rm NTN} \) & 1..20 \\
        TBS altitude [m] & \( h_{\rm TN} \) & 25 \\
        UE altitude [m] & \( h_{\rm UE} \) & 1.5 \\
        Total number of UEs & \(N_{\rm UE}\) & 570 \\
        Inter-site distance [m] & \(ISD\) & 500 \\
        Typical size of dense urban building [m] & \(W\) & 40.8 \\
        Typical dense urban street width [m] & \(S\) & 16.9 \\
    \end{tabular}
    \label{tab:param}
    \vspace{-1.5em}
\end{table}
\subsubsection{Network deployment}

Specifically, we adopted a hexagonal network layout consisting of \(19\) site locations separated by an inter-site distance \(ISD\).
In the \ac{TN} deployment, each site comprises three sectorized \acp{BS} with boresight orientations of \(30^\circ\), \(150^\circ\), and \(270^\circ\), resulting in a total of \(57\) coverage cells. 
We consider both \ac{4G} and \ac{5G} scenarios, where the \ac{BS} antennas are positioned at altitude \(h_{\rm TN}\).
In the \ac{NTN} deployment, \ac{5G} \acp{BS} are mounted on \acp{HAPS} at altitude \(h_{\rm NTN}\). 
Each \ac{HAPS}-based \ac{BS} is equipped with three circular-aperture reflector antennas, 
as specified in~\cite{3GPP38811}, creating an equivalent sectorized pattern.
Figure~\ref{fig:haps_deployment}a illustrates the considered \ac{NTN} deployment over a hexagonal grid, where the site locations are marked with blue dots. 
The green circles represent example coverage areas of the reflector antennas, while the red dots indicate the intersection of the antenna boresight with the ground surface.

\subsubsection{Frequency bands}

We assume that both terrestrial \ac{5G} and \ac{HAPS}-based deployments operate at the carrier frequency \(f_c^{\rm 5G}\) within the frequency band \(B_0^{\rm 5G}\). 
In contrast, the terrestrial \ac{4G} scenario operates at the carrier frequency \(f_c^{\rm 4G}\) within the frequency band \(B_0^{\rm 4G}\).
Full frequency reuse is assumed in all considered scenarios.

\subsubsection{User distribution}

For simplicity, 
we consider \(N_{\rm UE}\) outdoor single-antenna \acp{UE} positioned at a height \(h_{\rm UE}\) above ground level and uniformly distributed across all coverage cells. 
This assumption enables a controlled evaluation of network performance under balanced load conditions, without introducing additional variability due to non-uniform \ac{UE} spatial distributions.
In both terrestrial and non-terrestrial deployments, 
each \ac{UE} is associated with a single serving \ac{BS},
either ground-mounted in the \ac{TN} scenario or \ac{HAPS}-mounted in the \ac{NTN} scenario.

\subsection{NTN scenario adjustments}

In this subsection, we describe the additional modeling assumptions and adaptations required for the \ac{NTN} deployment scenario.

\subsubsection{Reflector antenna tilt}

In the \ac{NTN} deployment scenario,
the three reflector antennas serving the adjacent sectors of the same site are assumed to be co-located on a single \ac{HAPS}. 
Each antenna is oriented toward the ground in order to focus its beam on the intended coverage area.
To achieve the desired coverage geometry, 
the antenna must be tilted from the vertical direction by an angle \(\alpha\).
The specific value depends on the \ac{HAPS} altitude and is selected such that the antenna boresight intersects the center of the corresponding service area. 
Figure~\ref{fig:haps_deployment}b illustrates the antenna tilt configuration; here point \(A\) denotes the antenna vertical location on the \ac{HAPS}, the line \(AF\) represents the antenna boresight direction, 
while the segment \(NF\) lies on the ground surface and denotes the horizontal ground distance between points \(N\) and \(F\), and, finally, the angle \(HPBW\) denotes the half-power beamwidth of the reflector antenna.
From the deployment geometry, 
the horizontal distance between the site location and the center of the sector coverage area can be approximated as
\begin{equation}
NF = \frac{ISD}{3}.
\end{equation}

Accordingly, the required antenna tilt angle can be expressed as
\begin{equation}
\alpha = \arctan\!\left(\frac{ISD}{3h_{\rm NTN}}\right).
\end{equation}

Figure~\ref{fig:haps_deployment}b illustrates how the half-power beam footprint increases with altitude from 1~km to 4~km, assuming an aperture radius of 10 wavelengths.

\subsubsection{Channel models}

Generally, for each \ac{UE} $u$ and each cell $c$, the large-scale channel gain $\beta_{u,c}$ is defined as, 
\begin{equation}
    \beta_{u,c} = \rho_{u_c}\,\tau_{u_c}\, g_{u,c}\,\,, 
\end{equation}
where $\rho_{u_c}$ is the path loss gain,  $\tau_{u_c}$ is the shadowing gain and $g_{u,c}$ is the element antenna gain, computed from the \ac{3GPP} statistical channel models defined in~\cite{3GPP38901} for \ac{TN} and in~\cite{3GPP38811} for \ac{NTN} networks, respectively.
%
%
%
When considering the \ac{NTN} scenario, path loss gain $\rho_{u_c}^{\rm NTN}$ is modeled following ITU-R recommendations~\cite{ITU-R_P.618-14}  as follows, 
\begin{equation}
        \rho_{u_c}^{\rm NTN} = 1/{\rho_{u_c}^{\rm fspl} \, \rho_{u_c}^{\rm cl} \, \rho_{u_c}^{\rm ga} \, \rho_{u_c}^{\rm ra} \, \rho_{u_c}^{\rm ca} \, \rho_{u_c}^{\rm sa}}\,,
\end{equation}
where $\rho_{u_c}^{\rm fspl}$ is the free space propagation loss, $\rho_{u_c}^{\rm cl}$ is the clutter loss and $\rho_{u_c}^{\rm ga}$ $\rho_{u_c}^{\rm ra}$, $\rho_{u_c}^{\rm ca}$ $\rho_{u_c}^{\rm sa}$ are the  gaseous, rain, cloud and scintillation attenuation, respectively.

%
%
%
Additionally, for the \ac{NTN} deployment, the antenna element gain is replaced by the antenna reflector gain $g^r_{u,c}$ defined in~\cite{3GPP38811}:
\begin{equation}
    g^r_{u,c}(\theta') = 
\begin{cases}
1, & \theta' = 0 \\
4 \left| \dfrac{J_1(kr_m \sin \theta')}{kr_m \sin \theta'} \right|^2, & 0 < |\theta'| \leq 90^\circ
\end{cases}
\end{equation}
where $\theta'$ is the \ac{UE} zenith angle computed w.r.t. the normal direction of the reflector,
the  $J_1(\cdot)$ is the Bessel function of the first kind and first order, $k = 2\pi f/c $ is the wave number ($f$ is a carrier frequency, $c = 3 \times 10^8$ m/s), and $r_m$ is the physical radius of the circular aperture.

Then, the small-scale channel ${\bf h}_{u,c,k}$ for each \ac{PRB} \(k\), is modelled as a Rician fading channel and computed as,
\begin{equation}\label{eq:ComplexChannelRician}
    {\bf h}_{u,c,k} =
    \sqrt{\frac{K}{1+K}} \, {\bf h}^{\rm LOS}_{u,c,k}
    +
    \sqrt{\frac{1}{1+K}} \,  {\bf h}^{\rm NLOS}_{u,c,k} ,
\end{equation}
where $K$ denotes the Rician factor, ${\bf h}^{\rm NLOS}_{u,c,k}$ represents the \ac{NLoS} channel component modeled as Rayleigh fading, and ${\bf h}^{\rm LOS}_{u,c,k}$ is the \ac{LoS} component which follows a plane-wave approximation as defined in~\cite{3GPP38901}.

Finally, it is worth noting that the channel model in~\cite{3GPP38811} considers \ac{HAPS} location above 8 km. 
Meanwhile, placing \ac{HAPS}-mounted \acp{BS} at lower altitudes may increase the \ac{NLoS} probability due to geometric coverage constraints and greater interaction with ground-level urban clutter.
To address this issue,
we extended the channel model for \ac{NTN} scenario,
and adopted the \ac{LoS} probability calculation method designed by Sabour~\textit{et al.}~\cite{saboor2024geometry}. 
Particularly, the \ac{LoS} probability in the \ac{NTN} scenarios with \ac{HAPS} deployment altitude below 8 km is calculated by the following formula:
\begin{multline}
    \mathrm{PLoS}(\theta, S, W) = 
    \frac{S W}{A}\left( P^{R_1}_{\mathrm{LoS}}(\theta, S, W) + P^{R_2}_{\mathrm{LoS}}(\theta, S, W) \right) \\   
    + \frac{S^{2}}{A}\, P^{R_3}_{\mathrm{LoS}}(\theta, S, W),
    \label{eq:final_PLOS}
\end{multline} 
where \(\theta\) is an elevation angle,
\(P^{R_1}_{LoS}\) and \(P^{R_2}_{LoS}\) are \ac{LoS} probabilities for the \acp{UE} in street regions, while \(P^{R_3}_{LoS}\) is the \ac{LoS} probability for the users in the crossroad region,
\(W\) and \(S\) denote the typical size of the buildings and the street width correspondingly. 
The total streets and crossroad area can be calculated as \(A = 2SW + S^2\) while \(W\) and \(S\) are defined according to~\cite{ITU_R_P1410_6_2023}, 
based on the ratio of land covered by buildings to the total area and the mean number of buildings per unit area.
These parameters are pre-defined for suburban, urban, dense urban and high-rise urban scenarios, 
so we applied dense urban scenario parameters and calculated weighted average~\ac{LoS} probability for each user directly from the formula~\ref{eq:final_PLOS}.
The algorithm of \ac{LoS} probability calculation based on the formula~\ref{eq:final_PLOS}, 
including the calculation of \(P^{R_1}_{LoS}\), \(P^{R_2}_{LoS}\), and \(P^{R_3}_{LoS}\), 
is presented in detail in~\cite{saboor2024geometry},
while actual applied values of \(W\) and \(S\) are presented in the Table~\ref{tab:param}.

\subsection{Cell association, SINR and rate computation}

In both terrestrial and non-terrestrial deployments, 
each \ac{UE} is associated with the cell providing the highest \ac{RSRP}. 
For a given \ac{UE} \(u\) and candidate cell \(c\), the \ac{RSRP} is computed from the average received power of the corresponding reference signals, 
accounting for the transmit power, antenna gains, and the above channel model effects.
Then, the serving cell of \ac{UE} \(u\) is denoted by \(\hat{c}_u\).

Once the serving cell is selected,
the terrestrial downlink data \ac{SINR} of \ac{UE} \(u\) on \ac{MIMO} layer \(l\) and \ac{PRB} \(k\) is computed as
\begin{equation}
    \gamma_{u,l,k}^{\rm TN} =
    \frac{
        \beta^{\rm TN}_{u,\hat{c}_u}
        \left| \mathbf{h}_{u,\hat{c}_u,k}
        \mathbf{w}^{l}_{u,\hat{c}_u,k} \right|^2
        p^{l}_{u,\hat{c}_u,k}
    }{
        \sum\limits_{c \neq \hat{c}_u}
        \sum\limits_{u' \in \mathcal{U}_c}
        \sum\limits_{l'}
        \beta^{\rm TN}_{u,c,k}
        \left| \mathbf{h}_{u,c,k}
        \mathbf{w}^{l'}_{u',c,k} \right|^2
        p^{l'}_{u',c,k}
        + \sigma_k^2
    }
    \label{eq:SINR_Computation_TN}
\end{equation}
where \(\mathcal{U}_c\) denotes the set of \acp{UE} served by cell \(c\), 
\({\bf w}_{u,c,k}^{l}\) and \(p_{u,c,k}^{l}\) denote the precoding vector and transmit power allocated by cell \(c\) to \ac{UE} \(u\) on layer \(l\) and \ac{PRB} \(k\), respectively. 
The term \(\sigma_k^2\) represents the noise power on \ac{PRB} \(k\). 
Without loss of generality, 
each cell is assumed to uniformly distribute its transmit power across the allocated \acp{PRB} and beams.

Then, when considering the \ac{NTN} deployment, similarly, the downlink data \ac{SINR} of \ac{UE} \(u\) on \ac{PRB} \(k\) served by its serving cell  \(\hat{c}_u\), is computed as, 
\begin{equation}
    \gamma_{u,k}^{\rm NTN} =
    \frac{
    \beta^{\rm NTN}_{u,\hat{c}_u}
        \left| h_{u,\hat{c}_u,k} \right|^2
        p_{u,\hat{c}_u,k}
    }{
        \sum\limits_{c \neq \hat{c}_u}
        \sum\limits_{u' \in \mathcal{U}_c}
        \beta^{\rm NTN}_{u,c_u}
        \left| h_{u,c,k} \right|^2
        p_{u',c,k}
        + \sigma_k^2
    }
    \label{eq:SINR_Computation_NTN}
\end{equation}
where $\beta^{\rm NTN}$ denotes the large-scale channel gain for the \ac{HAPS}-based system, computed by accounting for the reflector antenna gain $g^r_{u,c}$. In this formulation, the reflector is modeled as a single antenna, and therefore the channel vector $\mathbf{h}_{u,c,k}$ reduces to a scalar and no antenna precoding is applied.

In both terrestrial and non-terrestrial deployments, 
the effective \ac{SINR} of \ac{UE} \(u\) on layer \(l\), 
denoted by $\tilde{\gamma}_{u,l}$, 
is then obtained from the set of per-\ac{PRB} \acp{SINR} \(\gamma_{u,l,k}\), using a mutual-information-based effective-\ac{SINR} mapping framework. 
This effective \ac{SINR} provides an overall representation of the quality experienced by the \ac{UE} over its allocated frequency resources.

Assuming round-robin scheduling and full-buffer traffic,
the achievable downlink rate of \ac{UE} \(u\) is computed as
\begin{equation}
    R_u = \sum_{l=1}^{L}
    \frac{
    N^{\rm PRB}_{\hat{c}_u} \, B^{\rm PRB}_{\hat{c}_u}
    }{
    N^{\rm UE}_{\hat{c}_u,l}
    }
    \log_2\!\left(1+\tilde{\gamma}_{u,l}\right),
    \label{eq:AchievableDataRate}
\end{equation}
where \(N^{\rm PRB}_{\hat{c}_u}\) is the total number of \acp{PRB} available at the serving cell \(\hat{c}_u\), 
\(B^{\rm PRB}_{\hat{c}_u}\) is the bandwidth of each \ac{PRB}, 
and \(N^{\rm UE}_{\hat{c}_u,l}\) is the number of \acp{UE} scheduled on \ac{MIMO} layer \(l\) of the serving cell.
Then, for the \ac{NTN} \ac{HAPS}-based holds $L=1$.


%% file: Sections/03_Results.tex
\section{Simulation Results}\label{sec:Results}

In this section, 
we present the main results of the performance analysis comparing the proposed \ac{HAPS}-based \ac{NTN} deployment with conventional terrestrial \ac{4G} and \ac{5G} cellular networks. 
The experiments were conducted using \textit{Giulia}, 
a system-level simulator calibrated according to \ac{3GPP} channel and deployment models to ensure realistic network-level performance evaluation. 
The simulation parameters used throughout the experiments are summarized in Table~\ref{tab:param}.

To ensure a fair comparison between terrestrial and \ac{HAPS}-based deployments, we assume that each sector in the \ac{5G} \ac{TN} operates on a single \ac{MIMO} layer and transmits one dominant beam toward the served coverage region. This configuration enables a consistent comparison between the \ac{HAPS} architecture and the terrestrial \ac{4G} and \ac{5G} baselines.
The investigation of more advanced multi-layer and massive \ac{MIMO} configurations is left for future work.

\subsection{Impact of HAPS Altitude and Reflector Aperture}

\begin{figure}[!tb]
    \centering
    \includegraphics[width=0.93\linewidth]{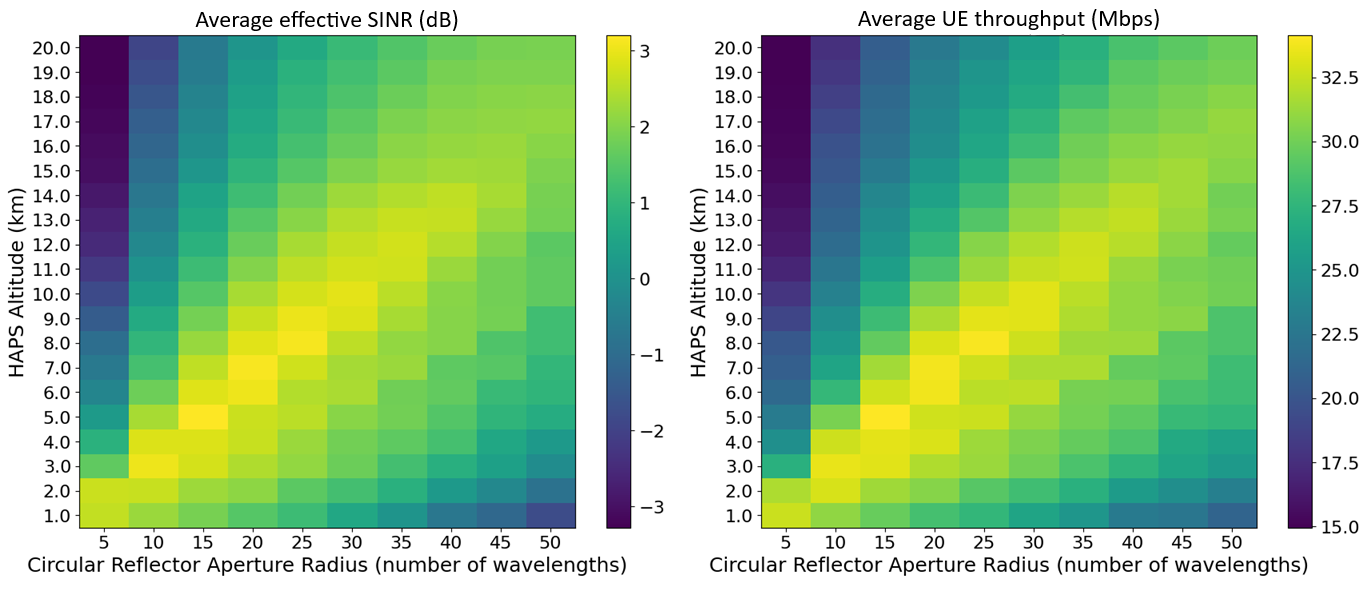}
    \caption{Average effective SINR (left) and UE throughput (right) of the HAPS-based deployment as functions of the platform altitude and reflector antenna aperture radius.}
    \label{fig:refl_performance}
\vspace{-1.5em}
\end{figure}

First, we analyze the performance of the \ac{HAPS}-based deployment across a wide range of platform altitudes (from 1\,km to 20\,km) and reflector aperture radii $r_{\lambda} = f_c^{\rm 5G}r_m/c$ measured in wavelengths (from 5 to 50). 
The resulting average effective \ac{SINR} at the \ac{UE} side and the corresponding average \ac{UE}  throughput are illustrated in Figure~\ref{fig:refl_performance} (left and right panels, respectively).
The figure highlights the joint impact of the platform altitude and reflector aperture on the overall network performance.
%
The results reveal a clear trade-off between the reflector aperture and the deployment altitude.
As the \ac{HAPS} altitude increases, the ground footprint of the transmitted beams expands (see Figure~\ref{fig:haps_deployment}b), resulting in stronger overlap between neighboring sectors and increased inter-cell interference.
To compensate for this effect, larger reflector apertures are required in order to generate narrower beams and limit the interference leakage toward adjacent cells.
Consequently, the optimal reflector aperture grows approximately linearly with altitude.
For example, at $h_{\rm NTN}$\,=\,5\,km, the highest average throughput (about 35\,Mbps) was observed with $r_{\lambda}$\,=\,15, while with $r_{\lambda}$\,=\,25 wavelengths the highest average throughput (about 34\,Mbps) was achieved at $h_{\rm NTN}$\,=\,8\,km.
Deviations from this operating region lead either to excessive interference (when beams are too wide) or to insufficient coverage (when beams are too narrow).
To demonstrate the impact of altitude and aperture deviations, in Figure~\ref{fig:reflector_vs_TN} we provide the performance comparison of the \ac{HAPS}-based configurations at $h_{\rm NTN}$\,=\,8\,km with $r_{\lambda}$ of 5, 10, 15, 30 and 50 wavelengths. For $r_{\lambda}$\,=\,25 wavelengths, the performance results at the altitudes of 4, 8, 12, 16, and 20\,km are also provided.
This observation indicates that the reflector configuration represents a key design parameter in \ac{HAPS}-based cellular deployments.

\begin{figure}[tb]
    \centering
    \includegraphics[width=1.01\linewidth]{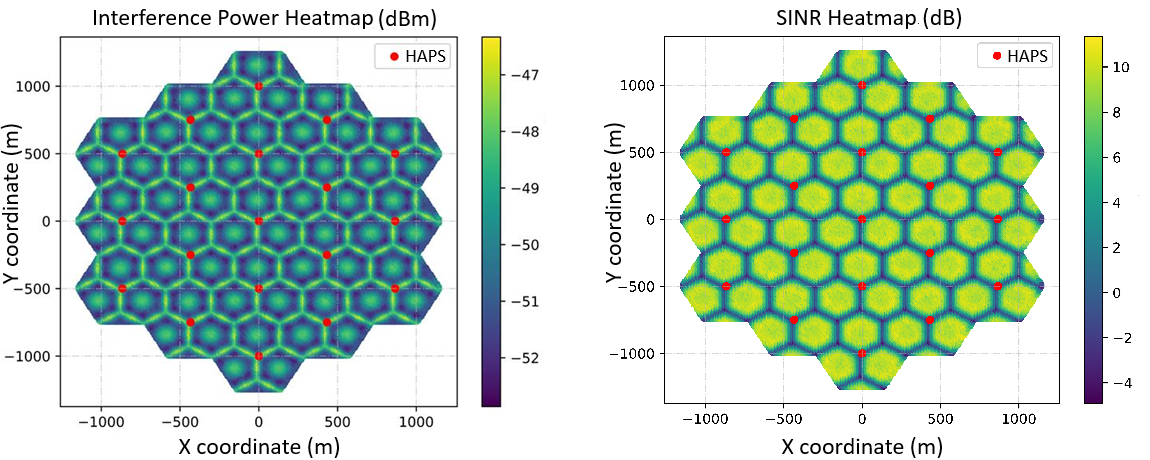}
    \caption{Spatial heatmaps of interfering power and SINR across the hexagonal deployment, considering $h_{\rm NTN}$ = 5\,km and $r_{\lambda}$ = 15.}
    \label{fig:multicell_heatmap}
    \vspace{-1.5em}
\end{figure}

\subsection{Performance Comparison with Terrestrial Deployments}

\begin{figure}[!tb]
    \centering
    \includegraphics[width=1\linewidth]{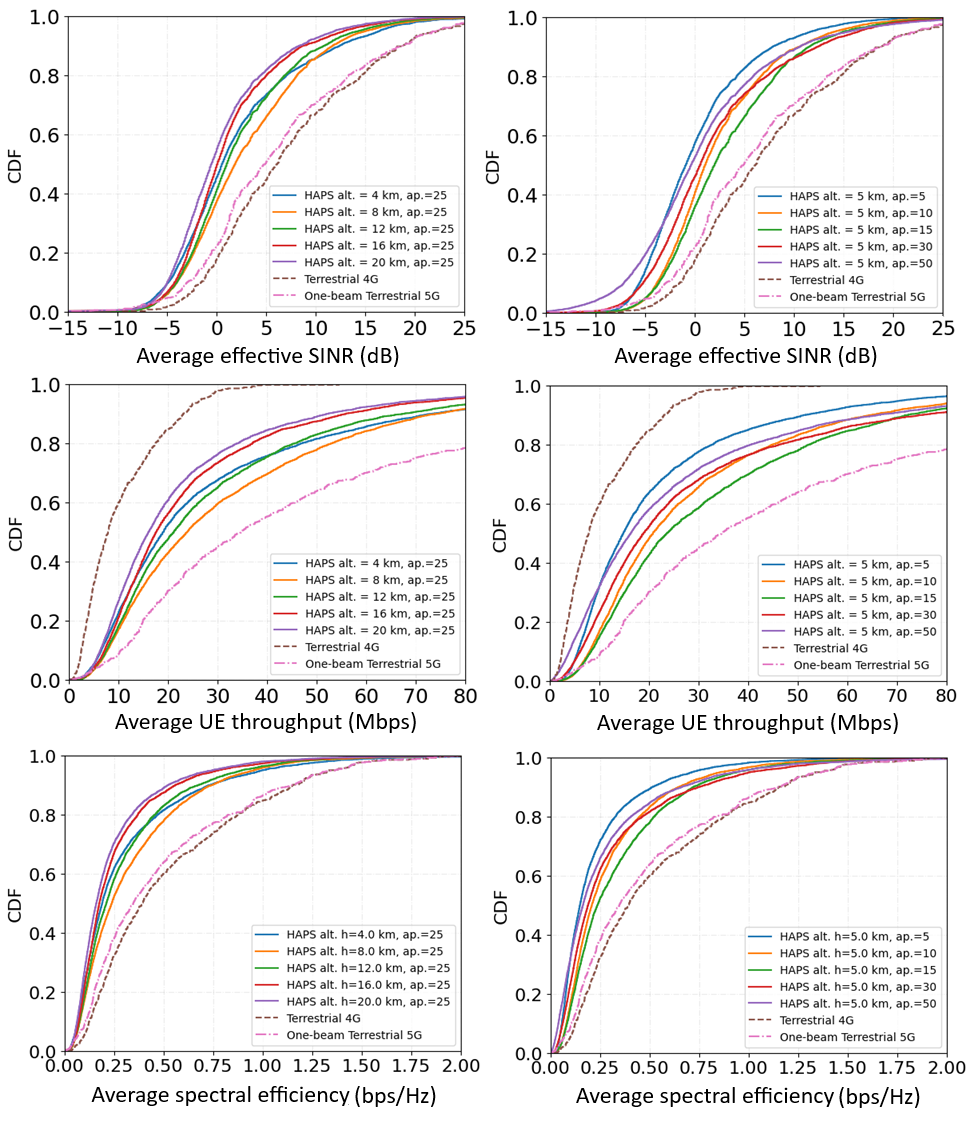}
    \caption{HAPS-based NTN vs. TN deployment: UE CDFs for SINR, throughput, and spectral efficiency.}
    \label{fig:reflector_vs_TN}
    \vspace{-1em}
\end{figure}

Next, we compare the performance of the optimal \ac{5G} \ac{HAPS}-based configurations with conventional terrestrial \ac{4G} and \ac{5G} deployments.
Figure~\ref{fig:reflector_vs_TN} shows the resulting average effective \ac{SINR} and average \ac{UE} throughput observed at the \ac{UE} side.

Under optimal reflector configurations ($h_{\rm NTN}$\,=\,8\,km, $r_{\lambda}$\,=\,25; $h_{\rm NTN}$\,=\,5\,km, $r_{\lambda}$\,=\,15), the median \ac{UE} throughput achieved by the \ac{5G} \ac{HAPS}-based deployment reaches 23.9-24.3\,Mbps, whereas the terrestrial \ac{5G} baseline achieves 33.5\,Mbps.
Meanwhile 5-percentile values for \ac{NTN} \ac{5G} scenarios reach 5.9-6.4\,Mbps, which are slightly higher than 5.2\,Mbps achieved by \ac{TN} \ac{5G}.
Although the \ac{HAPS} system benefits from favorable line-of-sight propagation conditions, the larger coverage footprint of each sector results in stronger inter-sector interference, which limits the achievable \ac{SINR}.
As expected, the terrestrial \ac{4G} deployment achieves lower throughput due to its smaller system bandwidth and less advanced transmission techniques.
Specifically, it reaches the median \ac{UE} throughput of 7.9\,Mbps and 5-percentile value of 1.9\,Mbps.

To provide a bandwidth-independent comparison, 
Figure~\ref{fig:reflector_vs_TN} also presents the average spectral efficiency per user for the considered deployments.
The spectral efficiency is obtained by normalizing the achieved throughput by the corresponding system bandwidth.
While the median spectral efficiency of the optimal \ac{5G} \ac{HAPS}-based configurations reaches approximately 0.24 bps/Hz per \ac{UE},
the terrestrial scenarios reach approximately 0.34\,bps/Hz for \ac{5G} configuration and 0.39\,bps/Hz for \ac{4G} configuration.
This difference confirms that the performance gap between terrestrial and \ac{5G} \ac{HAPS}-based systems is primarily caused by increased interference levels. 


\subsection{Interference Analysis in HAPS-Based Deployments}

To further investigate the origin of this performance gap,
we analyze the spatial distribution of the received signal quality and interference across the deployment area.
Figure~\ref{fig:multicell_heatmap} shows the heatmap of the average 
interfering received power across the hexagonal network layout.
It reveals that high-interference regions emerge near the center of each coverage sector which decrease the average effective \ac{SINR} up to 5\,dB in this area.
This effect arises from the overlap of the main lobes transmitted by neighboring sectors within the symmetric hexagonal deployment geometry.
Although the reflector antennas generate strong directional gain toward the intended coverage areas, the large altitude of the platform causes partial overlap of adjacent beams, which creates localized interference hotspots.
These interference regions partially offset the high useful signal power delivered by the main lobe of the reflector antenna.

Overall, the results indicate that the reflector-based \ac{HAPS} architecture is primarily constrained by interference coupling rather than by coverage limitations.
While the elevated platform provides strong line-of-sight propagation and wide-area coverage, the resulting beam overlap between neighboring sectors leads to an interference-limited operating regime in dense urban environments.
Consequently, even under optimized configurations, the effective \ac{SINR} achieved by the \ac{HAPS}-based deployment remains below that of terrestrial \ac{5G} systems operating with comparable bandwidth and numerology.

%% file: Sections/04_Conclusion.tex
\section{Conclusion}\label{sec:Conclusion}
In this work, we compared the performance of \ac{5G} \ac{NTN} enabled by \ac{HAPS}-based \acp{BS} equipped with tri-sector reflector antennas against conventional \ac{4G} \ac{TN} and \ac{5G} \ac{TN} deployments in a multi-cell dense urban environment.
A clear dependence on deployment altitude and reflector aperture is observed.
Simulation results and the following interference analysis demonstrated that overall system performance is primarily constrained by inter-cell interference.
Based on these findings and in order to close the performance gap against \ac{5G} \ac{TN}, our future work will adopt uniform planar array antennas combined with specifically designed beamforming algorithms for \ac{HAPS}-based \ac{NTN} deployments in dense urban scenarios.

%% file: journalAbbreviations.bib
@STRING{IEEE = {The Institute of Electrical and Electronics Engineers}}

@STRING{IEEE_J_COML       = "{IEEE} Commun. Lett."}

@STRING{IEEE_OJ_COMM       = "{IEEE} O.J. on Commun."}

@STRING{procictc = "Proc. IEEE Int. Conf. on Inf. and Comm. Tech. Conv. (ICTC)"}

@STRING{procvtc = "Proc. IEEE Veh. Tech. Conference (VTC)"}


%% file: refs.bib
@misc{3GPP38901,
 author = {{Study on channel model for frequencies from 0.5 to 100 GHz, 3GPP TR38.901}},
 note = {v.18.0},
 month = mar,
 year = {2024}
}

@misc{3GPP38811,
 author = {{Technical Specification Group Radio Access Network; Study on New Radio (NR) to support non-terrestrial networks, 3GPP TR38.811}},
 note = {v.15.4},
 month = sep,
 year = {2020}
}

@ARTICLE{saboor2024geometry,
  author={Saboor, Abdul and Vinogradov, Evgenii and Cui, Zhuangzhuang and Al-Hourani, Akram and Pollin, Sofie},
  journal=IEEE_OJ_COMM, 
  title={A Geometry-Based Modelling Approach for the Line-of-Sight Probability in {UAV} Communications}, 
  year={2024},
  volume={5},
  number={},
  pages={364-378},
  month=dec,
  doi={10.1109/OJCOMS.2023.3341627}
}

@article{svistunov2025bridgingearthspacesurvey,
      title={Bridging Earth and Space: A Survey on HAPS for Non-Terrestrial Networks}, 
      author={G. Svistunov and A. Akhtarshenas and D. López-Pérez and M. Giordani and G. Geraci and H. Yanikomeroglu},
      year={2025},
      eprint={2510.19731},
      archivePrefix={arXiv},
      primaryClass={eess.SY},
      journal={arXiv:2510.19731},
      url={https://arxiv.org/abs/2510.19731}, 
}

@ARTICLE{anicho2021multihaps,
  author={Anicho, Ogbonnaya and Charlesworth, Philip and  Baicher, Gurvinder and Nagar, Atulya},
  journal={IJICTA}, 
  title={{Multi-HAPS network implementation within 3GPP’s NTN framework for 5G and beyond}}, 
  year={2021},
  volume={7},
  number={1},
  pages={7-12},
  doi={10.17972/ijicta20217152}
}

@INPROCEEDINGS{xing2021architecture,
  author={Xing, Yunchou and Hsieh, Frank and Ghosh, Amitava and Rappaport, Theodore S.},
  booktitle=procvtc,
  title={High Altitude Platform Stations ({HAPS}): Architecture and System Performance}, 
  year={2021},
  volume={},
  number={},
  pages={1-6},
  month=jun,
  doi={10.1109/VTC2021-Spring51267.2021.9448899}
}

@INPROCEEDINGS{lee2023performance,
  author={Lee, Jaeyeol and Kim, Tae-Yoon and Kim, Jae-Hyun},
  booktitle=procictc,
  title={Performance Analysis of Multiple High Altitude Platform Stations Cellular Network Coverage},
  year={2023},
  volume={},
  number={},
  pages={1242-1244},
  month=oct,
  doi={10.1109/ICTC58733.2023.10393398}
}

@ARTICLE{shibata2020system,
  author={Shibata, Yohei and Kanazawa, Noboru and Konishi, Mitsukuni and Hoshino, Kenji and Ohta, Yoshichika and Nagate, Atsushi},
  journal={IEEE Access}, 
  title={System Design of Gigabit {HAPS} Mobile Communications}, 
  year={2020},
  volume={8},
  number={},
  pages={157995-158007},
  month=aug,
  doi={10.1109/ACCESS.2020.3019820}
}

@ARTICLE{shamsabadi2024enchancing,
  author={Shamsabadi, Afsoon Alidadi and Yadav, Animesh and Yanikomeroglu, Halim},
  journal=IEEE_J_COML,
  title={Enhancing Next-Generation Urban Connectivity: Is the Integrated {HAPS}-Terrestrial Network a Solution?}, 
  year={2024},
  volume={28},
  number={5},
  pages={1112-1116},
  month=feb,
  doi={10.1109/LCOMM.2024.3370698}
}

@ARTICLE{shibata2024haps,
  author={Shibata, Yohei and Takabatake, Wataru and Hoshino, Kenji and Nagate, Atsushi and Ohtsuki, Tomoaki},
  journal={IEEE Access}, 
  title={HAPS Cell Design Method for Coverage Extension Considering Coexistence on Terrestrial Mobile Networks}, 
  year={2024},
  volume={12},
  number={},
  pages={55506-55520},
  doi={10.1109/ACCESS.2024.3390116}
}

@INPROCEEDINGS{liu2022interference,
  author={Liu, Wenjia and Hou, Xiaolin and Chen, Lan and Hokazono, Yuki and Zhao, Jinming},
  booktitle=procvtc,
  title={Interference Coordination Method for Integrated {HAPS}-Terrestrial Networks}, 
  year={2022},
  volume={},
  number={},
  pages={1-6},
  month=jun,
  doi={10.1109/VTC2022-Spring54318.2022.9860546}
}

@misc{ericsson2025reportNov,
  author       = {{Ericsson}},
  title        = {{Ericsson Mobility Report, November 2025}},
  institution  = {Ericsson},
  year         = {2025},
  month        = nov,
  url          = {https://www.ericsson.com/en/reports-and-papers/mobility-report/reports/november-2025},
  note         = {\url{https://www.ericsson.com/en/reports-and-papers/mobility-report/reports/november-2025} Accessed: 29-Jan-2026}
}

@techreport{ITU-R_P.618-14,
  author      = {{ITU-R}},
  title       = {Propagation data and prediction methods required for the design of Earth-space telecommunication systems},
  institution = {International Telecommunication Union},
  type        = {Recommendation},
  number      = {P.618-14},
  year        = {2021},
  month       = {December},
  url         = {https://www.itu.int}
}

@techreport{ITU_R_P1410_6_2023,
  author       = {{ITU-R}},
  title        = {Propagation data and prediction methods required for the design of terrestrial broadband radio access systems operating in a frequency range from 3 GHz to 60 {{GHz}}},
  institution = {International Telecommunication Union},
  type        = {Recommendation},
  number       = {P.1410-6},
  year         = {2023},
  month        = {Aug},
  url          = {https://www.itu.int/dms_pubrec/itu-r/rec/p/R-REC-P.1410-6-202308-I!!PDF-E.pdf}
}

@techreport{HAPSAlliance2026WhitePaper,
  author       = {{HAPS Alliance}},
  title        = {HAPS for 6G: Expanding the Connectivity Landscape},
  institution  = {HAPS Alliance},
  year         = {2026},
  month        = mar,
  url          = {https://hapsalliance.org/wp-content/uploads/formidable/12/HAPSAlliance_6G_WhitePaper_MAR26.pdf},
  note         = {White paper; accessed: 13-Apr-2026}
}
